# Practical Synchronization Waveform for Massive Machine-Type-Communications

M. Wang, and JJ Zhang

*Abstract*—This paper proposes a practical synchronization waveform that is resilient to frequency error for machine-type-communications (MTC) with applications in massive Internet-of-Things (IoT). Mathematical properties of the waveform are derived, which are keys to addressing the practical issues. In particular, it is shown that this type of waveform is *asymptotically optimal* in the presence of a frequency error, in the sense that its asymptotic performance is the same as the optimal matched-filter detector that is free of frequency error. This asymptotic property enables optimization of the waveform under the constraints imposed by an application. It is also shown that such optimized waveform comes in pairs, which facilitates the formation of a new waveform capable of frequency error estimation and timing refinement at the receiver in addition to the resilience to receiver frequency errors.

## I. Introduction

Machine-type-communications (MTC) is a type of wireless communications that support fully automatic data generation, exchange, processing, and actuation among intelligent machines, without or with low intervention of humans including utilities, sensing, health care, manufacturing, and transportation [1], [2]. Whereas massive MTC (mMTC) characterized by simultaneous support of a massive number of MTC devices is becoming the prominent communication paradigm for a wide range of emerging smart services with a typical application in the massive Internet of Things (mIoT) market where devices sending bits of information to other machines, servers, clouds, or humans account for a much larger proportion in wireless communication applications.

Although cellular networks such as GSM and LTE have long been used for MTC in various IoT applications, its capability to support mMTC is rather limited. Nevertheless, this is not surprising since cellular technologies were developed for "human-type" communications (HTC) in the first place. To provide a solution that are built on top of the traditional cellular network for mMTC, the Third Generation Partnership Project (3GPP) dedicated an immense effort during LTE Release 13 to develop a new radio access technology known as Narrow-Band Internet of Things (NB-IoT) [3] as part of the long term evolution process towards a more versatile universal communication technology. As such, NB-IoT inherits most functionality from the legacy LTE system, such as the transmission frame structure and the data transmission waveform (i.e., OFDM). The most noticeable changes are probably the reduced minimum system bandwidth from 1.4 MHz to 180 kHz, mainly for exploiting the refarmed GSM spectrum that are channelized 200 kHz per carrier, and the redesigned synchronization waveform for better serving mIoT use cases that are commonly characterized by low-cost, extended coverage, and the unique short burst, long sleep transmission pattern [4].

Synchronization is the first and arguably the most challenging step of MTC in mIoT applications due to the largest time and frequency uncertainties present in transceivers. Because of the low-cost nature of an mIoT device, a local oscillator of the device may suffer from a large frequency error that creates an offset in carrier frequency between the incoming signal and the receiver. For instance, the initial frequency error of the local oscillator can be as large as 20 ppm for an mMTC device [5]. Moreover, the short-burst nature of an mMTC transmission makes synchronization of each transmission a more dominant factor in the overall transmission efficiency as compared to that in HTC; and the prolonged sleep duration (to save battery) causes the local oscillator to drift away from the default frequency. These unique characteristics of mMTC further burden the synchronization process. Yet, data transmissions can proceed only after time and frequency synchronization is established as required by data transmission waveforms. Since the purposes of the waveform used for synchronization and the ones for data transmission (i.e., the multiple access waveforms) are different, the design requirements for these two types of waveforms are very different as well.

Compared to the data transmission or multiple access waveform [6]-[13], the synchronization waveform for mMTC so far has received much less attention in 5G technology development. Recently in [14], a general synchronization waveform resilient to frequency error is derived. It is shown that the effect of a frequency error on the well-known matched filter-based detector of such type of waveform is simply a time shift of the detection peak position without incurring a significant loss in detection energy. However, for use as a practical synchronization waveform, issues like (1) waveform parameter selection to meet application-specific requirements; and (2) the capability for frequency error estimation and timing refinement that is *essential* to a synchronization signal still remain to be resolved. This paper proposes a practical synchronization waveform that solves these important issues while retaining the frequency error resilience property.

In Section II, we briefly review the derivation of the synchronization waveform proposed in [14]. In Section III, we deal with the first issue, i.e., the selection of waveform parameters, detailing the optimization of the waveform under the constraints imposed by a practical application, which is exemplified by NB-IoT. Key mathematical properties of the waveform are derived along the way as needed. In Section IV, we shift our focus to the second issue, presenting the means for frequency error estimation and timing refinement exploiting the unique properties of this type of waveforms derived. Finally, Section V concludes this paper.

## II. Frequency Offset Resilient Waveform

Assuming the baseband synchronization waveform is represented as $x(t)$, the local copy of this synchronization waveform of a matched-filter based detector (also known as the cross-correlation based detector) in effect becomes $x(t)e^{j2\pi\Delta f t}$, in the presence of a frequency offset $\Delta f$.

The cross-correlation function between the incoming synchronization signal $x(t)$ and the local copy can be written as

$$\gamma(\tau,\Delta f) = \frac{1}{\mathcal{E}} \int_{-T/2}^{T/2} x(t) e^{j2\pi\Delta f t} x^*(t-\tau) dt, \quad (1)$$

where $\mathcal{E} = \int_{-T/2}^{T/2} |x(t)|^2 dt$.

In the absence of a frequency offset, i.e., $\Delta f = 0$, the maximum output of the correlator happens when $\tau = 0$,

$$\gamma(\tau=0, \Delta f=0) = 1, \quad (2)$$

providing the optimal detection performance as asserted by the well-known matched-filter detection theory. While in the presence of a frequency offset, $\Delta f \neq 0$ (unknown to the receiver), the phase ramping component, $e^{j2\pi\Delta f t}$, introduced by the frequency offset effectively creates a mismatch between the incoming synchronization signal and the local copy of the waveform, breaking the very premise of the optimality of a matched-filter detector, thereby inevitably resulting in a loss in detection energy, i.e.,

$$\gamma(\tau=0, \Delta f \neq 0) < 1, \quad (3)$$

and ergo a loss in detection performance. This phenomenon is well-documented in the literature [14]-[19], and becomes prominent in mIoT when the frequency offset is likely large – typically to an extent that the resulting mismatch totally fails a matched filter-based detector.

In NB-IoT, this issue is dealt with by repeating a waveform multiple times consecutively in time, and a differential correlator between repeated signals (also called an autocorrelator) is employed at the receiver to suppress the phase ramping effect. Specifically, the waveform is repeated 11 times to form a synchronization signal with a total length of ~780 μs. Although very effective (in removing phase ramping), this type of autocorrelator-based detector is not optimal, causing at least 3 dB degradation at its best (as SNR $\to +\infty$) with respect to the optimal matched-filter detector (without frequency offset). The degradation quickly increases as SNR deteriorates, e.g., close to 5 dB degradation at an SNR of -5 dB, and ~10 dB at -10 dB. This behavior, known as the "noise amplification" phenomenon, could be problematic in a low SNR scenario which is not uncommon in mIoT deployments.

In search for a solution such that the optimality of a matched filter detector can be maximally retained under frequency offset, in [14], matched-filter (i.e., cross-correlator) based detection with frequency offset is re-examined using the Cauchy-Schwartz inequality, which indicates that

$$|\gamma(\tau,\Delta f)| \leq \frac{1}{\mathcal{E}} \sqrt{\int_{-T/2}^{T/2} |x(t) e^{j2\pi\Delta f t}|^2 dt} \cdot \sqrt{\int_{-T/2}^{T/2} |x^*(t-\tau)|^2 dt} . \quad (4)$$

That is,

$$|\gamma(\tau,\Delta f)| \leq 1, \quad \forall \Delta f \text{ and } \forall x(t), \quad (5)$$

with equality if and only if

$$x(t) e^{j2\pi\Delta f t} = C \cdot x(t-\tau), \quad (6)$$

where $C \in \mathbb{C}$ is a non-zero constant.

The significance of this result is obvious in that it claims the *mathematical* existence of such a waveform that attains the optimality of a matched filter even in the presence of a frequency offset between the received signal and the local waveform, as long as condition (6) is met.

It can be shown that the waveform that satisfies condition (6) is of the following general form

$$\tilde{x}(t) = e^{j\pi(\alpha t^2 + \beta t)}, \quad \alpha,\beta \in \mathbb{R} . \quad (7)$$

The matched filter output energy based on this type of waveform transforms the frequency error of the detector, $\Delta f$, into a time shift away from the original position by an amount of

$$\hat{\tau} = -\alpha^{-1} \Delta f . \quad (8)$$

## III. Prototype Waveform and Optimality

In the previous section, we have briefly reviewed the derivation of a general waveform in [14] that has the capability of converting a frequency offset between the transmitted signal and the detector into a time offset. However, the question regarding how to take advantage of this property of this type of waveform to build a *practical* synchronization waveform remains to be answered since to be a practical synchronization waveform, it must allow us to select waveform parameters to satisfy the application requirements; and more importantly, estimate the frequency and time offset/error at the receiver, which is, after all, one of the *essential* functions of a synchronization signal, nevertheless unavailable in [14]. In this section, we first form a prototype waveform, and show its asymptotic optimality. We then utilize this property for optimizing the waveform subject to practical constraints. The analytical results from this section pave the way to Section IV.

### A. Waveform Constraints

In the mathematical treatment of Section II, it is implicitly assumed that $\tilde{x}(t)$ extends beyond the correlation interval,

$[-T/2, T/2]$. In practice, a synchronization waveform is time-bounded within length $T$, i.e.,

$$\hat{x}_{\alpha,\beta}(t) \triangleq \begin{cases} \tilde{x}(t), & -T/2 \leq t \leq T/2 \\ 0, & \text{otherwise} \end{cases}. \quad (9)$$

This practical form of $\tilde{x}(t)$ defined in (7) is the prototype for the following analysis, and serves as the building block for creating the ultimate synchronization waveform in Section IV.

As such, when a frequency error causes a shift of the correlation peak from $\tau=0$ to $\hat{\tau}$, it creates a time offset or misalignment, $\hat{\tau}$, between the received waveform and the local one, causing that only partial signal energy can be detected, which consequently results in a loss in detection peak energy. Therefore, the effect of this time misalignment on the detection energy (and ergo the detection performance) needs to be taken into account in practical designs.

It is apparent that the misalignment, $\hat{\tau}$, between the incoming signal and the local waveform must be less than the waveform length, $T$, in order for the correlator to output nonzero detection energy, i.e.,

$$|\hat{\tau}| < T. \quad (10)$$

It is not difficult to find that the loss in detection energy due to a time offset $\hat{\tau}$ is

$$\ell(\hat{\tau}) = \left(1 - T^{-1}|\hat{\tau}|\right)^2. \quad (11)$$

From (10) it is clear that

$$0 < \ell(\hat{\tau}) < 1 \quad (12)$$

in linear, or

$$\ell(\hat{\tau}) < 0 \quad (13)$$

in dB, for $\hat{\tau} \neq 0$ (i.e., $\Delta f \neq 0$).

Hence, the practical form of the mathematical waveform $\tilde{x}(t)$, i.e., the prototype $\hat{x}_{\alpha,\beta}(t)$ given in (9), no longer attains the maximum detection energy of an actual matched filter. In fact, it is $-\ell(\hat{\tau})$ dB away from the optimal. The selection of the waveform parameters, i.e., $\alpha$ and $\beta$, therefore needs to minimize this deficit, or maximize $\ell(\hat{\tau})$, i.e.,

$$\langle \dot{\alpha}, \dot{\beta} \rangle = \arg\max_{\alpha, \beta \in \mathbb{R}} \ell(\hat{\tau}). \quad (14)$$

From (8) this deficit is found to be a sole function of parameter $\alpha$ (not a function of $\beta$). Substituting (8) into (11) follows that

$$\ell(\alpha) = \left(1 - T^{-1}|\Delta f|/|\alpha|\right)^2 \\ = \left(1 - \mu/|\alpha|\right)^2, \quad (15)$$

where

$$\mu \triangleq T^{-1} \cdot |\Delta f| > 0. \quad (16)$$

Clearly, $\ell(\alpha) = 1$ (or 0 dB), i.e., no loss with respect to the optimal, when $\mu = 0$ or $\Delta f = 0$.

Equation (14) then becomes

$$\dot{\alpha} = \arg\max_{\alpha \in \mathbb{R}} \ell(\alpha), \quad (17)$$

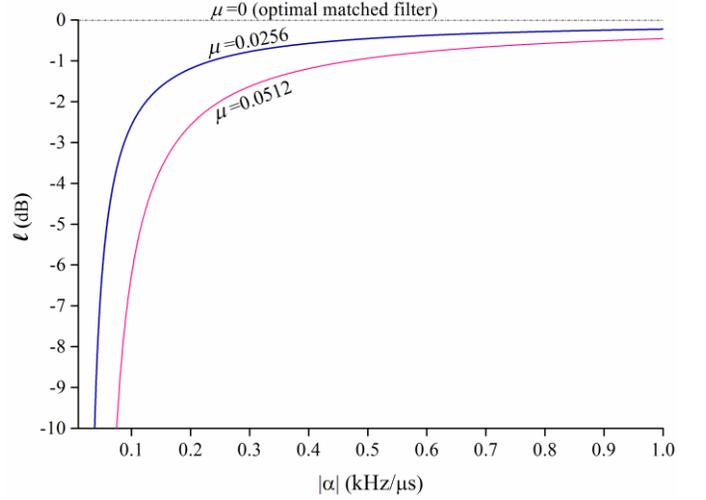

Fig. 1. Illustration of the asymptotic property of the prototype waveform $\hat{x}(t)$ in the presence of a non-zero frequency error, where the cases for $\mu = 780^{-1} \times 20 = 0.0256$ and $\mu = 390^{-1} \times 20 = 0.0512$ (kHz/μs) are shown.

which is equivalent to

$$\dot{\alpha} = \arg\max_{\alpha \in \mathbb{R}} |\alpha|. \quad (18)$$

From (8), (10) and (16), it is clear that

$$|\alpha| > \mu. \quad (19)$$

It is not difficult to see from (15) that, for nonzero $\mu$, i.e., $\Delta f \neq 0$,

$$\ell(\alpha) \to 1^- \text{ or } 0^- \text{ dB}, \quad (20)$$

as $|\alpha| \to +\infty$. We thus have the following proposition:

*Proposition 1*: Waveform $\hat{x}(t)$ defined in (9) *asymptotically* attains the optimality of a matched-filter detector, in the presence of a frequency error. The rate of convergence is determined by $\mu$ defined in (16), which is irrelevant of SNR.

This asymptotic behavior of $\hat{x}_{\alpha,\beta}(t)$ is graphically shown in Fig. 1, where the dotted line is the asymptote, i.e., $\ell(\alpha) = 0$ (dB) for $\mu = 0$ (i.e., $\Delta f = 0$), as promised by an actual matched-filter detector that has the full knowledge of the input signal frequency, i.e., zero frequency error, and the cases with $\mu \neq 0$ ($\Delta f \neq 0$): $\mu = 0.0256$ and $0.0512$ (kHz/μs), corresponding to $T = 780$ μs and $T/2 = 390$ μs, respectively, given $|\Delta f| = 1$ GHz × 20 ppm = 20 kHz. Here we have borrowed the NB-IoT parameters, the primary synchronization signal length ($T = 780$ μs), and maximum frequency error ($|\Delta f| = 20$ ppm), as example.

It is now evident that the design of the prototype waveform $\hat{x}_{\alpha,\beta}(t)$ optimized for a particular application becomes *maximization* of the magnitude of the waveform parameter $\alpha$, subject to the specific constraint imposed by the application.

## B. Frequency Error constraint

From (8) and (10), it is clear that

$$|\alpha^{-1}\Delta f| < T. \quad (21)$$

This implies that for a maximum supported frequency error requirement, $\Delta f_{max}$, the selection of $\alpha$ must satisfy

$$S_1: \quad |\alpha| > |\Delta f_{max}|/T. \quad (22)$$

That is to say, for a given waveform length, the larger the maximum frequency error that a practical system is designed to tolerate, the greater the magnitude of $\alpha$ is required.

For NB-IoT, the maximum supported frequency error is $|\Delta f_{max}| = 20$ ppm, corresponding to ~20 kHz at 1 GHz carrier frequency. Hence the solution to (22) is

$$S_1: \quad \langle \alpha_1, \beta_1 \rangle \in \mathcal{C}_1^\dagger \quad (23)$$

where

$$\mathcal{C}_1^\dagger \triangleq \{\langle \alpha, \beta \rangle \big| |\alpha| > 0.0256\} \quad (24)$$

with $T = 780 \, \mu s$.

For the same frequency error tolerance but a shorter waveform length, e.g., $T/2 = 390 \, \mu s$, the corresponding solution is

$$\mathcal{C}_1^\dagger \triangleq \{\langle \alpha, \beta \rangle \big| |\alpha| > 0.0512\}. \quad (25)$$

## C. Spectral Constraints

Both (18) and (22) demand a large $|\alpha|$. Nevertheless, in practice, the selection of $\alpha$ cannot be arbitrarily large and is often limited by the application-specific constraints. A typical example of such constraints is the spectral requirements, e.g., the maximum occupied bandwidth restriction, and the adjacent channel leakage ratio (ACLR) requirement [20][21]. The occupied bandwidth is the width of a frequency band such that, below the lower and above the upper frequency limits, the powers emitted are each equal to a specified percentage $\sigma/2$ (e.g., $\sigma = 1\%$) of the total transmitted power, while ACLR is the ratio of the power centred on the assigned channel frequency to that centred on an adjacent channel frequency.

Therefore, the optimization of the waveform in (18) can be reformulated as

$$\dot{\alpha} = \arg\max_{S_1 \cap S_2 \cap S_3} |\alpha|, \quad (26)$$

where $S_1$ is the constraint from the supported maximum frequency error requirement given in (22), $S_2$ is the occupied bandwidth requirement, and $S_3$ the ACLR requirement.

In detail, $S_2$ requires that a specified percent of the waveform energy be confined within a given bandwidth $W \in \mathbb{R}^+$, i.e.,

$$S_2: \quad \frac{1}{T}\int_{-W/2}^{W/2} \hat{\mathcal{X}}_{\alpha,\beta}(f) df \geq 1 - \sigma, \quad (27)$$

where

$$\hat{\mathcal{X}}_{\alpha,\beta}(f) \triangleq \left| \int_{-T/2}^{T/2} \hat{x}_{\alpha,\beta}(t) e^{-j2\pi ft} dt \right|^2, \quad \forall f \in \mathbb{R}, \quad (28)$$

is the power spectrum of waveform $\hat{x}(t)$, and noting that

$$\mathcal{E} = \int_{-T/2}^{T/2} |\hat{x}_{\alpha,\beta}(t)|^2 dt = T.$$

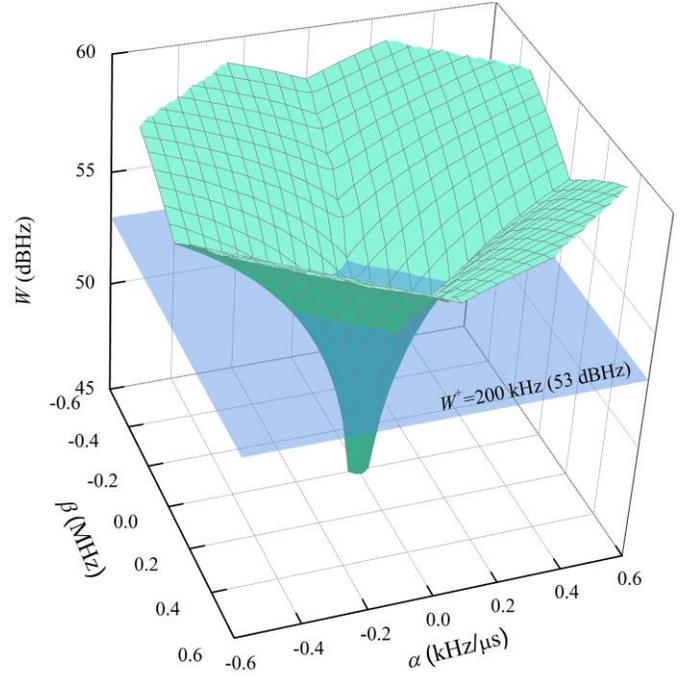

(a)

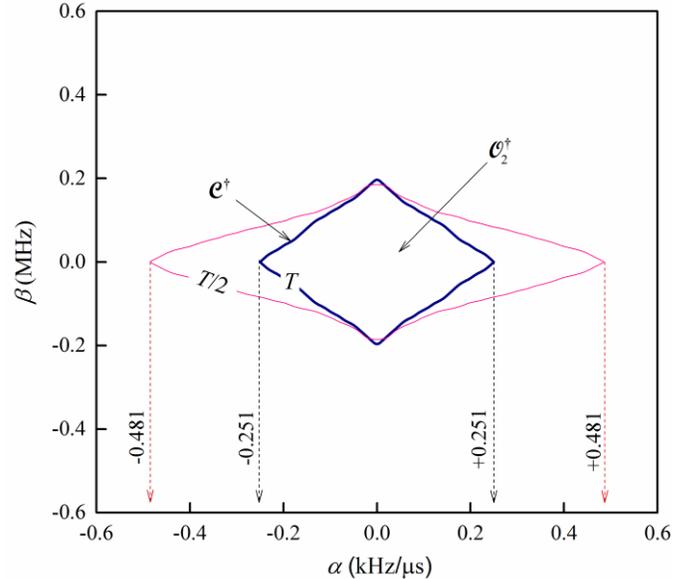

(b)

Fig. 2 Graphical illustration of occupied bandwidth constraint, $S_2$, on the parameters $\alpha$ and $\beta$: (a) the occupied bandwidth surface $\mathcal{W}_\sigma$ for $T = 780 \, \mu s$; and (b) the contour $\mathcal{C}^\dagger$ formed by $\langle \alpha, \beta \rangle$'s whose $W^\dagger = 200$ kHz (53 dBHz) for $T = 780 \, \mu s$. $T/2 = 390 \, \mu s$ is also plotted.

Using NB-IoT as example, $W = 200$ kHz, $\sigma = 1\%$, and $T = 780 \, \mu s$ [22].

To see the implication of $S_2$ on the waveform parameters $\alpha$ and $\beta$, let us first look at (27) with equality, from which the occupied bandwidth, $W$, of waveform $\hat{x}_{\alpha,\beta}(t)$ can be plotted as a function of $\alpha$ and $\beta$, which produces a surface as illustrated in Fig. 2 (a). Every point on this surface is a pair of $\alpha$ and $\beta$ values, $\langle \alpha, \beta \rangle$, associated with a $W \in \mathbb{R}^+$ that corresponds to a particular bandwidth, within which $1-\sigma = 99\%$ of the energy of $\hat{x}_{\alpha,\beta}(t)$ is confined. This surface, henceforth referred to as *the occupied bandwidth surface*, is denoted as $\mathcal{W}_\sigma(\alpha,\beta)$, and mathematically represented as

$$\mathcal{W}_\sigma(\alpha,\beta) \triangleq \left\{ W \in \mathbb{R}^+ \,\middle|\, \frac{1}{T}\int_{-W/2}^{W/2} \hat{\mathcal{X}}_{\alpha,\beta}(f)df = 1-\sigma \right\}. \quad (29)$$

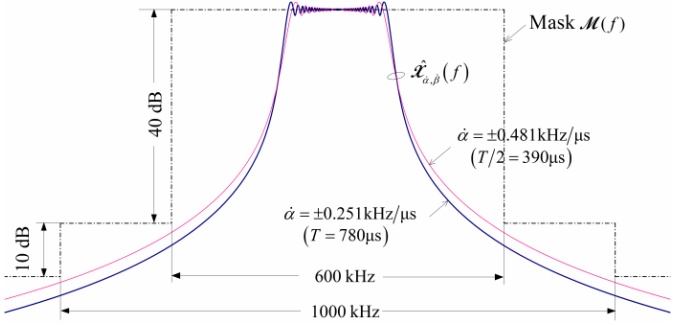

Fig. 3. The ACLR requirement: the NB-IoT spectrum mask $\mathcal{M}(f)$ and the $\hat{\mathcal{X}}_{\alpha,\beta}(f)$ of the optimal waveform.

Among these $\alpha$-$\beta$ pairs on $\mathcal{W}_\sigma$, we are particularly interested in the ones that correspond to a given $W$, $W^\dagger$, i.e.,

$$\mathcal{C}^\dagger \triangleq \left\{ \langle \alpha, \beta \rangle \,\middle|\, \mathcal{W}_\sigma(\alpha,\beta) = W^\dagger \right\}, \quad (30)$$

which forms a closed symmetric contour $\mathcal{C}^\dagger$ on the $\alpha$-$\beta$ plane as illustrated in Fig. 2 (b) where $W^\dagger = 200$ kHz (or 53 dBHz) and $\sigma = 99\%$, which can be viewed as the intersection between the occupied bandwidth surface $\mathcal{W}_\sigma(\alpha,\beta)$ and the $W^\dagger = 200$ kHz plane [see Fig. 2 (a)].

In fact, it can be further shown that any $\alpha$-$\beta$ pair that falls within the area enclosed by $\mathcal{C}^\dagger$ satisfies (27). If we denote this enclosure (including $\mathcal{C}^\dagger$) as $\mathcal{O}_2^\dagger$, we can conclude that $S_2$ in (27) is satisfied by $\forall \langle \alpha, \beta \rangle \in \mathcal{O}_2^\dagger$. Consequently, $\hat{\mathcal{X}}_{\alpha,\beta}(f)$ with parameters $\alpha$ and $\beta$ selected from any point in $\mathcal{O}_2^\dagger$ meets the maximum occupied bandwidth requirement, $S_2$.

The occupied bandwidth requirement in (27) thus implies

$$S_2 : \langle \alpha_2, \beta_2 \rangle \in \mathcal{O}_2^\dagger, \quad (31)$$

as graphically shown in Fig. 2 (b).

$S_3$, the ACLR restriction requires that

$$S_3 : \hat{\mathcal{X}}_{\alpha,\beta}(f) \le \mathcal{M}(f), \quad \forall f \in \mathbb{R}. \quad (32)$$

where $\mathcal{M}(f)$ is the spectral mask.

The ACLR from NB-IoT indicates that for adjacent channel with 300 kHz offset is $-40$ dBc and with 500 kHz is $-50$ dBc, which can be represented as

$$\mathcal{M}(f) = \begin{cases} -40 \text{ (dBc)} & 300 \text{ (kHz)} \le |f| < 500 \text{ (kHz)} \\ -50 \text{ (dBc)} & |f| \ge 500 \text{ (kHz)} \end{cases} \quad (33)$$

as shown in Fig. 3, where $f$ is the frequency offset with respect to the carrier frequency. It is not surprising to see the relaxed ACLR requirement since the adjacent channel interference from NB-IoT is suppressed by the 1/12 frequency reuse plan for GSM, recalling that NB-IoT mainly targets the refarmed GSM bands. Nevertheless, the solution to (32) can be found to be

$$S_3 : \langle \alpha_3, \beta_3 \rangle \in \mathcal{O}_3^\dagger, \quad (34)$$

which is graphically shown in Fig. 4.

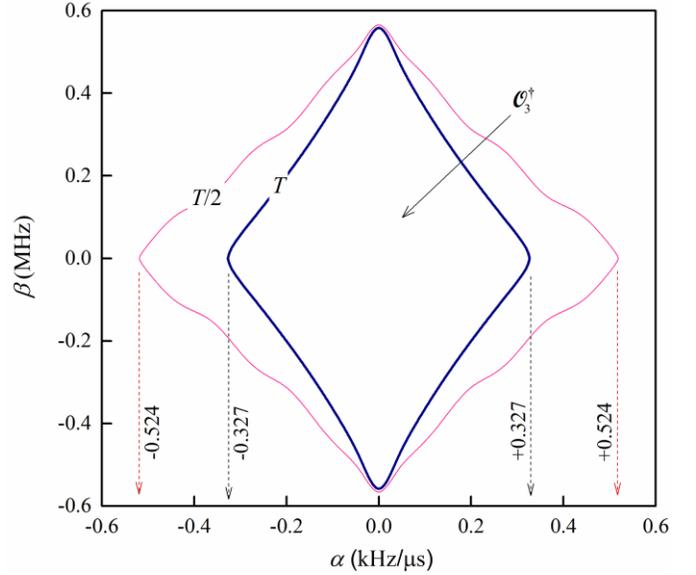

Fig. 4. Graphical illustration of the ACLR restriction, $S_3$, on waveform parameters $\alpha$ and $\beta$, for $T = 780$ μs as well as $T/2 = 390$ μs.

Now we are ready to find the solution to (26) by combining (23), (31), and (34),

$$\begin{aligned} \langle \dot{\alpha}, \dot{\beta} \rangle &= \arg\max_{S_1 \cap S_2 \cap S_3} |\alpha| \\ &= \arg\max_{\langle \alpha,\beta \rangle \in \mathcal{O}_1^\dagger \cap \mathcal{O}_2^\dagger \cap \mathcal{O}_3^\dagger} |\alpha| \\ &= \langle \pm 0.251, 0 \rangle. \end{aligned} \quad (35)$$

for $T = 780$ μs.

The solution in (35) gives the optimal parameters in the sense that the synchronization waveform, $\hat{x}_{\alpha,\beta}(t)$, parameterized by $\langle \dot{\alpha}, \dot{\beta} \rangle$ has the least performance loss (with respect to the optimal matched-filter) among all the waveforms in $\{\hat{x}_{\alpha,\beta}(t) | \alpha, \beta \in \mathbb{R}\}$, constrained by the maximum frequency error tolerance, maximum occupied bandwidth, and ACLR spectrum mask. Waveform $\hat{x}_{\dot{\alpha},\dot{\beta}}(t)$ is hence referred to as the *optimal waveform* in the same sense. The power spectrum $\hat{\mathcal{X}}_{\dot{\alpha},\dot{\beta}}(f)$ of the optimal waveform $\hat{x}_{\dot{\alpha},\dot{\beta}}(t)$ is plotted in Fig. 3.

From (11) or Fig. 1, the corresponding degradations of the optimal waveform parameterized by (35) at various magnitudes of frequency errors is plotted in Fig. 5. It is seen that the degradation is less than 1 dB at a maximum frequency error of 20 kHz, and dwindles to 0 dB as the frequency error becomes zero, which is significantly less than a differential detector which suffers much greater than 3 dB degradation, even without frequency errors.

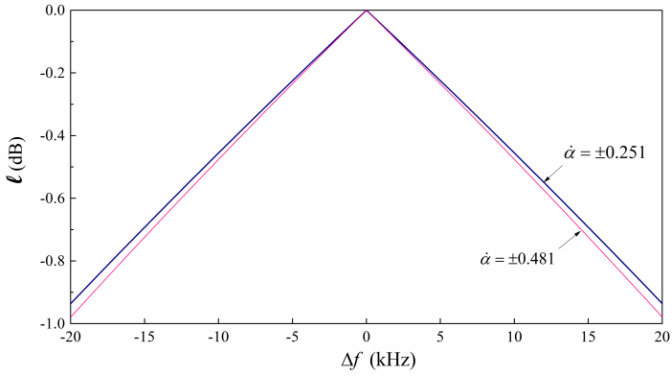

Fig. 5. Plot of the performance degradation (relative to the optimal matched-filter, 0dB) of the optimal waveform $\hat{x}_{\dot{\alpha},0}(t)$ against the frequency error $(\Delta f)$ ranging from $-20$ kHz to $+20$ kHz.

It is noted that the optimal value of $\alpha$ is a function of the time duration $T$ of the waveform. For instance, following the same optimization procedure, the optimal $\alpha$ for $T/2 = 390$ µs can be found to be

$$\dot{\alpha} = \arg\max_{S_1 \cap S_2 \cap S_3} |\alpha| = \pm 0.481 \ (\text{kHz/µs}). \tag{36}$$

## IV. PRACTICAL WAVEFORM

So far we have shown that the prototype waveform defined in (9) possesses certain useful mathematical properties that not only provide robust detection performance against frequency errors but also facilitate optimization to meet the application requirements. However, in addition to the primary role of detecting the presence and timing of a system, another essential function of a synchronization signal is to provide frequency synchronization. To this end, the frequency error $\Delta f$ needs to be obtained after signal $\hat{x}_{\alpha,\beta}(t)$ is detected. From (8), $\Delta f$ is linearly related to $\hat{\tau}$, the time shift of the detection peak from the *actual* timing position, however also *unknown* to the receiver. It is thus clear that, in its original form, the prototype waveform *cannot* be used as a practical synchronization signal.

To solve this dilemma, we need the following property, which is already seen from the solutions to (26), and is generally true.

*Proposition 2*: The optimal solutions to (26) come in *symmetric* pairs, i.e., if $\langle \dot{\alpha}, \dot{\beta} \rangle$ is the solution to (26), then $\langle -\dot{\alpha}, \dot{\beta} \rangle$ is also a solution.

This is because the constraint, $S_i$ ($i = 1, 2, 3$), is symmetric in terms of $\alpha$ and $\beta$. Consequently, $\mathcal{O}_i^\dagger$ is symmetric, i.e.,

(1) symmetric about $\alpha = 0$, i.e.,

$$\mathcal{O}_i^\dagger(-\alpha, \beta) = \mathcal{O}_i^\dagger(\alpha, \beta); \tag{37}$$

(2) symmetric about $\beta = 0$, i.e.,

$$\mathcal{O}_i^\dagger(\alpha, -\beta) = \mathcal{O}_i^\dagger(\alpha, \beta); \tag{38}$$

and

(3) symmetric about $\langle \alpha = 0, \beta = 0 \rangle$, i.e.,

$$\mathcal{O}_i^\dagger(-\alpha, -\beta) = \mathcal{O}_i^\dagger(\alpha, \beta). \tag{39}$$

This symmetry property is straightforward for $i = 1$, while for $i = 2$, and 3 it is a direct result from the fact that $\hat{\mathcal{X}}_{\alpha,\beta}(f)$ is symmetric, i.e.,

$$\hat{\mathcal{X}}_{-\alpha,\beta}(f) = \left| \int_{-T/2}^{T/2} \hat{x}_{-\alpha,\beta}(t) e^{-j2\pi ft} dt \right|^2$$
$$= \left( \int_{-T/2}^{T/2} e^{j\pi(-\alpha t^2 + \beta t)} e^{-j2\pi ft} dt \right) \left( \int_{-T/2}^{T/2} e^{j\pi(-\alpha t^2 + \beta t)} e^{-j2\pi ft} dt \right)^*. \tag{40}$$

Substituting $-t$ with $\varsigma$ results in

$$\hat{\mathcal{X}}_{-\alpha,\beta}(f)$$
$$= \left( \int_{-T/2}^{T/2} e^{j\pi(\alpha\varsigma^2 + \beta\varsigma)} e^{-j2\pi f\varsigma} d\varsigma \right) \left( \int_{-T/2}^{T/2} e^{j\pi(\alpha\varsigma^2 + \beta\varsigma)} e^{-j2\pi f\varsigma} d\varsigma \right)^* \tag{41}$$
$$= \hat{\mathcal{X}}_{\alpha,\beta}(f),$$

and (37) follows.

Similarly, it can be shown that

$$\hat{\mathcal{X}}_{\alpha,-\beta}(f) = \hat{\mathcal{X}}_{\alpha,\beta}(-f), \tag{42}$$

which leads to (38) since

$$\frac{1}{T} \int_{-W^\dagger/2}^{W^\dagger/2} \hat{\mathcal{X}}_{\alpha,\beta}(f) df \overset{\xi \triangleq -f}{=} \frac{1}{T} \int_{-W^\dagger/2}^{W^\dagger/2} \hat{\mathcal{X}}_{\alpha,\beta}(-\xi) d\xi$$
$$= \frac{1}{T} \int_{-W^\dagger/2}^{W^\dagger/2} \hat{\mathcal{X}}_{\alpha,-\beta}(f) df, \tag{43}$$

whereas (39) directly follows from (37) and (38).

The following proposition is a direct extension of Proposition 2.

*Proposition 3*: The optimal waveforms come in pairs, meaning that if $\langle \dot{\alpha}, \dot{\beta} \rangle$ parameterizes an optimal waveform,

$$\hat{x}_{\dot{\alpha},\dot{\beta}}(t) = \begin{cases} e^{j\pi(\dot{\alpha}t^2 + \dot{\beta}t)}, & -T/2 \leq t \leq T/2 \\ 0, & \text{otherwise} \end{cases}, \tag{44}$$

then $\langle -\dot{\alpha}, \dot{\beta} \rangle$ also yields an optimal waveform,

$$\hat{x}_{-\dot{\alpha},\dot{\beta}}(t) = \begin{cases} e^{j\pi(-\dot{\alpha}t^2 + \dot{\beta}t)}, & -T/2 \leq t \leq T/2 \\ 0, & \text{otherwise} \end{cases}. \tag{45}$$

Together, (44) and (45) constitute the *optimal waveform pair*.

Following the same example from the previous section, waveforms

$$\hat{x}_{\dot{\alpha},0}(t), \tag{46}$$

and

$$\hat{x}_{-\dot{\alpha},0}(t) = \hat{x}_{\dot{\alpha},0}^*(t), \tag{47}$$

where $\dot{\alpha} = 0.251$ (kHz/µs) with $T = 780$ µs, form an optimal pair that are co-conjugate.

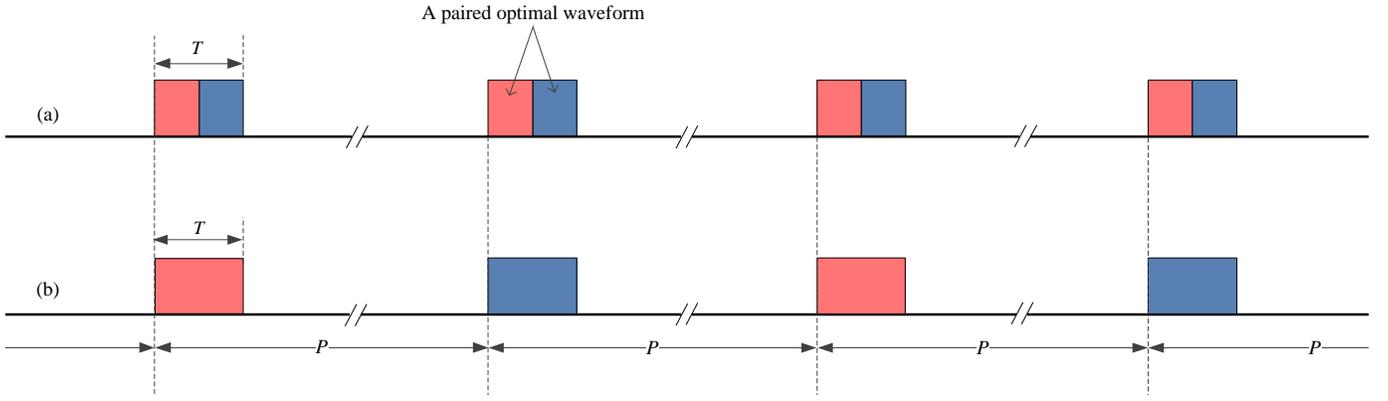

Fig. 6. Illustration of the paired optimal waveforms based on Proposition 3 that enables frequency error estimation, where red denotes $\hat{x}_{\dot\alpha,0}(t)$ and blue $\hat{x}_{-\dot\alpha,0}(t)$.

It is evident that the optimal waveform pair not only provide the same performance but also produce the same time shift, only in opposite directions. That is to say, a frequency error $\Delta f$ causes the output of the matched filter to $\hat{x}_{-\dot\alpha,\dot\beta}(t)$ to shift by

$$\hat{\tau} = \dot\alpha^{-1}\Delta f \tag{48}$$

in response to an incoming signal $\hat{x}_{-\dot\alpha,\dot\beta}(t)$, as opposed to

$$\hat{\tau} = -\dot\alpha^{-1}\Delta f \tag{49}$$

in response to $\hat{x}_{\dot\alpha,\dot\beta}(t)$ from its corresponding matched filter [cf. (8)].

This property leads to a simple means to obtain the time shift (and then the actual timing and the frequency error) by constructing a composite waveform composed of $\hat{x}_{\dot\alpha,0}(t)$ followed by its counterpart $\hat{x}_{-\dot\alpha,0}(t)$, i.e.,

$$\bar{x}(t) \triangleq \begin{cases} \hat{x}_{\dot\alpha,0}(t), & -T/4 \le t < T/4 \\ \hat{x}_{-\dot\alpha,0}(t-T/2), & T/4 \le t < 3T/4 \end{cases} \tag{50}$$

as depicted in Fig. 6 (a), where the length of $\bar{x}(t)$ is $T$, $P$ is the transmission period, and $\hat{x}_{\dot\alpha,0}(t)$ has length $T/2 = 390$ μs with $\dot\alpha = 0.481$ (kHz/μs). For a given $\Delta f$, this gives rise to two correlation peaks, one of which is located at

$$t_1 = \hat{\tau}_1 \tag{51}$$

in response to $\hat{x}_{\dot\alpha,0}(t)$ at the output of a matched filter to $\hat{x}_{\dot\alpha,0}(t)$, where

$$\hat{\tau}_1 = -\dot\alpha^{-1}\Delta f \tag{52}$$

according to (8), and the other at

$$t_2 = T/2 + \hat{\tau}_2 \tag{53}$$

in response to $\hat{x}_{-\dot\alpha,0}(t)$ at the output of another matched filter to $\hat{x}_{-\dot\alpha,0}(t)$, where

$$\hat{\tau}_2 = \dot\alpha^{-1}\Delta f \tag{54}$$

They are separated by a distance of $t_2 - t_1$ or

$$\hat{d} = T/2 + (\hat{\tau}_2 - \hat{\tau}_1) = T/2 + 2\cdot(\dot\alpha^{-1}\Delta f). \tag{55}$$

Fig. 7 is the sample outputs of the paired-correlator detector in response to the composite synchronization waveform, where (a) is the case with a frequency error of 20 kHz, (b) 0, and (c) −20 kHz.

Solving (55) for $\Delta f$ immediately yields

$$\Delta f = \frac{\dot\alpha}{2}\left(\hat{d} - \frac{T}{2}\right). \tag{56}$$

Since $\hat{d}$ can be directly measured after detection, $\Delta f$ can then be calculated. Apparently, the distance between the two detection peaks is $\hat{d} = \frac{T}{2}$, in the absence of a frequency error.

The composite waveform enables the estimation of a frequency error, and yet has an efficient detection implementation by taking advantage of the fact that most of the computations can be shared between the paired correlators due to the co-conjugate nature.

It is also possible to transmit the paired optimal waveforms $P$ sec apart, as illustrated in Fig. 6 (b), where they are transmitted alternatively in different transmission opportunities. In this case, the detector output peaks are separated further apart, and the corresponding frequency error is calculated as

$$\Delta f = \frac{\dot\alpha}{2}(\hat{d} - P). \tag{57}$$

With the knowledge of the frequency error, the timing error due to the time shift attributable to $\Delta f$ is readily available from (8). A correction can then be made to the detected timing to remove the effect of the frequency error on the detected timing.

Fig. 8 plots the cumulated error distribution function (CDF) of the estimated receiver frequency error via numerical simulations, where waveform in Fig. 6 (a) with parameter pair, $\langle \pm\dot\alpha, \dot\beta \rangle = \langle \pm 0.481, 0 \rangle$ [see (36)], is constructed and transmitted by a base station with power $p = 43$ dBm. Between two transmissions, random data are also transmitted to mimic real life situations.

The maximum coupling loss between the transmitter and the receiver of an mIoT device is 144 dB plus 20 dB additional penetration loss (for deployment deep inside a building, e.g., basement). The total propagation loss is thus $\Delta = 164$ dB. The receiver sensitivity is then

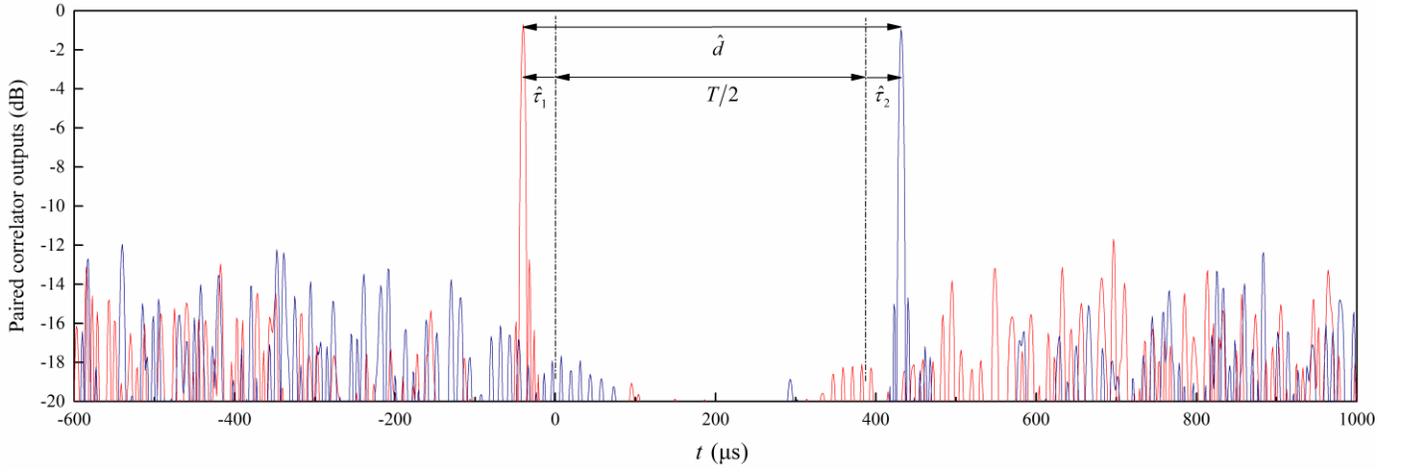

(a) $\Delta f = 20$ kHz

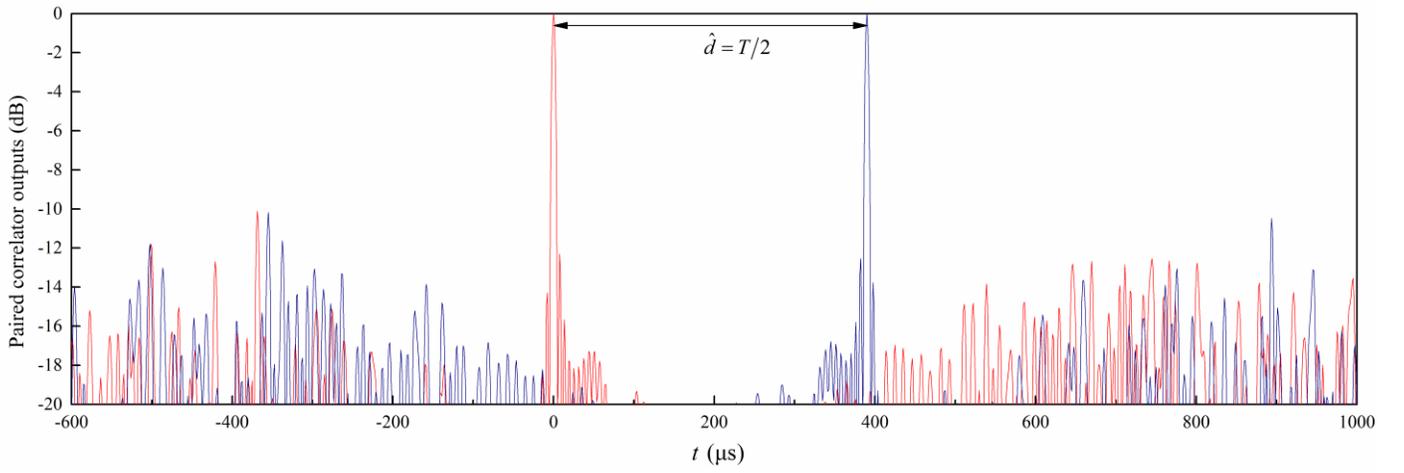

(b) $\Delta f = 0$

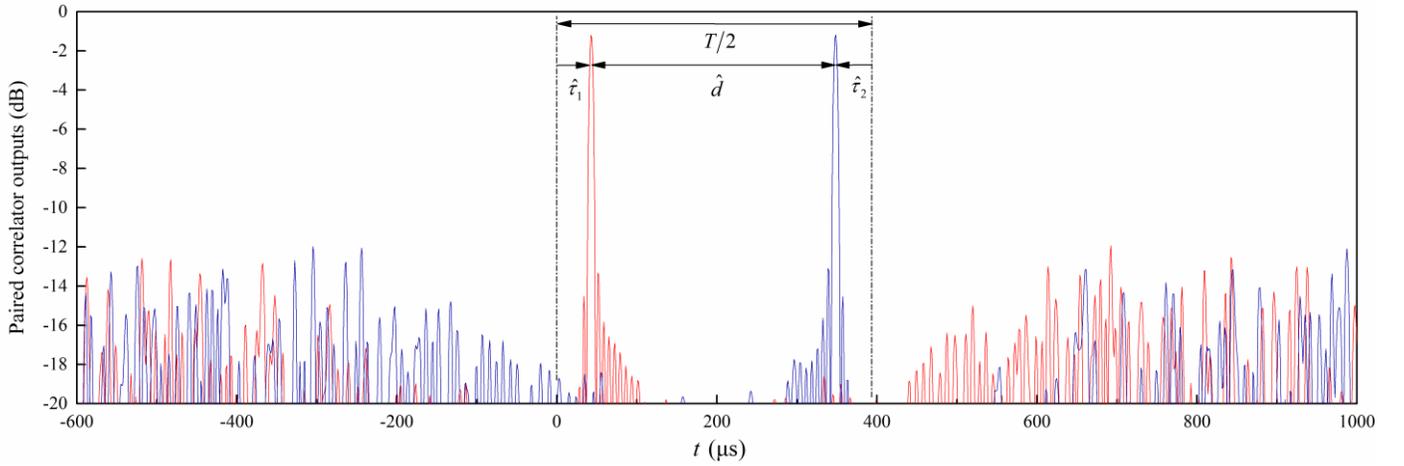

(c) $\Delta f = -20$ kHz

Fig. 7 Sample outputs of the paired correlators in response to an incoming sequence of $\bar{x}(t)$ in (50) with $\dot{\alpha}=0.481$ (kHz/μs) and data streams, resulting in paired peaks separated by $\hat{d}$ : (a) $\Delta f = 20$ kHz, (b) $\Delta f = 0$ , and (c) $\Delta f = -20$ kHz.

$$\rho = p - \Delta$$
$$= 43 \text{ dBm} - 164 \text{ dB}. \quad (58)$$
$$= -121 \text{ dBm}$$

The received signal is filtered with a 200 kHz bandwidth filter operating at a sample rate of $4 \times 200$ kHz. The noise figure of the mIoT receiver is $\xi = 5$ dB. The noise power at the receiver is thus

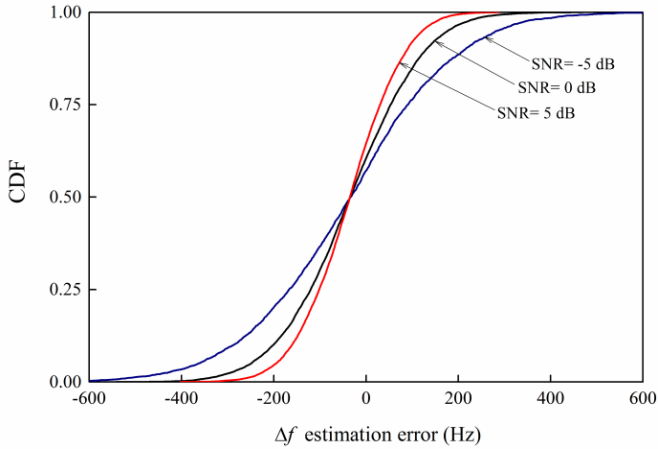

Fig. 8 Frequency error $(\Delta f)$ estimation error CDF with receiver signal SNR levels of 5, 0, and -5 dB. The frequency error is uniformly distributed in the range from -20 kHz to +20 kHz.

$$N = W + N_0 + \xi$$
$$= 53 \text{ dBHz} - 174 \text{ dBm/Hz} + 5 \text{ dB} \quad (59)$$
$$= -116 \text{ dBm},$$

where $N_0$ is the noise power spectral density (i.e., $-174$ dBm/Hz). The corresponding signal SNR at the receiver is

$$\eta = \rho - N$$
$$= -121 \text{ dBm} - (-116 \text{ dBm}), \quad (60)$$
$$= -5 \text{ dB}$$

which is the minimum SNR that an mIoT receiver operates at.

A frequency error $\Delta f$, uniformly distributed in the range from $-20$ kHz to $+20$ kHz, is introduced into the incoming signal, which then passes through a filter of bandwidth 200 kHz at a sampling rate of 1600 kHz. The detector employs a pair of matched filters to the paired waveforms, and *jointly* detects the incoming paired signals. The frequency error is estimated using (55) after the paired signals are successfully detected. The evaluation is performed at SNR levels of 5, 0, and -5 dB. It is observed from Fig. 8 that about 95 percent of $\Delta f$ estimation errors fall within 400 Hz, corresponding to $\hat{\tau}$ estimation error of less than 2 µs.

## V. CONCLUSION

In this paper, we propose a practical design of the frequency-error-resilient synchronization waveform for massive MTC. We derive and exploit the key mathematical properties of the prototype waveform for waveform optimization, frequency error estimation, as well as timing refinement. The design is exemplified by a specific practical application, i.e., LTE NB-IoT. We show that the practical form of this waveform is asymptotically optimal, in the sense that its asymptotic detection energy in the presence of frequency error is the same as an optimal matched filter which has full knowledge of the input signal frequency (i.e., free of frequency error). Based on this property, the practical design problem boils down to maximization of the waveform parameter, i.e., $\alpha$, under the constraints present in the application. We further show that the optimal parameter of this type of waveform comes in symmetric pairs, which facilitates the creation of a unique synchronization waveform consisting of paired optimal waveforms for determining the frequency error and refined timing at the receiver.

REFERENCES

[1] E. Dutkiewicz, X. Costa-Perez, and I. Z. Kovacs, "Massive machine-type communications," *IEEE Network*, vol. 31, no. 6, pp. 6-7, Nov. 2017.

[2] M. Wang, W. Yang, J. Zou, B. Ren, M. Hua, J. Zhang, X. You, "Cellular machine-type communications: Physical challenges and solutions," *IEEE Wireless Commun.*, vol. 23, no. 2, pp. 108-117, April 2016.

[3] 3rd Generation Partnership Project Technical Specification Group Radio Access Network Evolved Universal Terrestrial Radio Access (E-UTRA) Physical channels and modulation (Release 13), TS 36.211 ver. 13.2.0, June 2016.

[4] From GSM to LTE-Advanced Pro and 5G: An Introduction to Mobile Networks and Mobile Broadband, Martin Sauter, John Wiley & Sons, Inc., 111 River Street, Hoboken, NJ 07030, USA.

[5] Y.-P. E. Wang, X. Lin, A. Adhikary, A. Grovlen, Y. Sui, Y. Blankenship, J. Bergman, and H. S. Razaghi, "A primer on 3GPP narrowband Internet of Things," *IEEE Commun. Mag.*, vol. 55, no. 3, pp. 117-123, Mar. 2017.

[6] G. Berardinelli, K. I. Pedersen, T. B. Sørensen, and P. Mogensen, "Generalized DFT-Spread-OFDM as 5G waveform," *IEEE Commun. Mag.*, vol. 54, no. 11, pp. 99-105, Nov. 2016.

[7] X. Zhang, L. Chen, J. Qiu, and J. Abdoli, "On the waveform for 5G," *IEEE Commun. Mag.*, vol. 54, no. 11, pp. 74-80, Nov. 2016.

[8] A. A. Zaidi, R. Baldemair, H. Tullberg, H. Björkegren, L. Sundström, J. Medbo, C. Kilinc, and I. D. Silva, "Waveform and numerology to support 5G services and requirements," *IEEE Commun. Mag.*, vol. 54, no. 11, pp. 90-98, Nov. 2016.

[9] J. Yli-Kaakinen, T. Levanen, S. Valkonen, K. Pajukoski, J. Pirskanen, M. Renfors, and M. Valkama, "Efficient Fast-Convolution-Based waveform processing for 5G physical layer," *IEEE J. Sel. Areas Commun.*, vol. 35, no. 6, pp. 1309-1326, Jun. 2017.

[10] P. Guan, D. Wu, T. Tian, J. Zhou, X. Zhang, L. Gu, A. Benjebbour, M. Iwabuchi, and Y. Kishiyama, "5G field trials: OFDM-Based waveforms and mixed numerologies," *IEEE J. Sel. Areas Commun.*, vol. 35, no. 6, pp. 1234-1243, Jun. 2017.

[11] Y. Liu, X. Chen, Z. Zhong, B. Ai, D. Miao, Z. Zhao, J. Sun, Y. Teng, and H. Guan, "Waveform design for 5G networks: Analysis and comparison," *IEEE Access*, vol. 5, pp. 19282-19292, Jan. 2017.

[12] B. Farhang-Boroujeny and H. Moradi, "OFDM inspired waveforms for 5G," *IEEE Commun. Surveys Tuts.*, vol. 18, no. 4, pp. 2474-2492, 4th Quart., 2016.

[13] J. M. Hamamreh and H. Arslan, "Secure orthogonal transform division multiplexing (OTDM) waveform for 5G and beyond," *IEEE Commun. Lett.*, vol. 21, no. 5, pp. 1191-1194, May 2017.

[14] J. Zhang, M. Wang, M. Hua, W. Yang, and X. You, "Robust synchronization waveform design for massive IoT," *IEEE Trans. on Wireless Commun.*, vol. 16, no. 11, pp. 7551-7559, Nov. 2017.

[15] H. Kroll, M. Korb, B. Weber, S. Willi, and Q. Huang, "Maximum-likelihood detection for energy-efficient timing acquisition in NB-IoT," in *Proc. IEEE Wireless Communications and Networking Conference Workshops* (WCNCW), San Francisco, USA, Mar. 2017, pp. 1-5.

[16] M. Hua, M. Wang, W. Yang, X. You, F. Shu, J. Wang, W. Sheng, and Q. Chen, "Analysis of the frequency offset effect on random access signals," *IEEE Trans. Commun.*, vol. 61, no. 11, pp. 4728-4740, Nov. 2013.

[17] M. Hua, M. Wang, K. W. Yang, and K. J. Zou, "Analysis of the frequency offset effect on Zadoff–Chu sequence timing performance,"

*IEEE Trans. Commun.*, vol. 62, no. 11, pp. 4024-4039, Nov. 2014.

[18] H. Puska and H. Saarnisaari, "Matched filter time and frequency synchronization method for OFDM systems using PN-sequence preambles," in *IEEE 18th International Symposium on Personal, Indoor and Mobile Radio Communications*, Athens, 2007, pp. 1-5.

[19] F. Tufvsson, O. Edfors, and M. Faulkner, "Time and frequency synchronization for OFDM using PN-sequence preambles," in *Proc. IEEE Vehicular Technology Conf.*, vol. 4, Sep. 1999, pp. 2203-2207.

[20] ITU-R recommendation SM.328: "Spectra and bandwidth of emissions".

[21] ITU-R recommendation SM.329: "Unwanted emissions in the spurious domain".

[22] *Third Generation Partnership Program (3GPP)*: Evolved Universal Terrestrial Radio Access Base Station (BS) radio transmission and reception, Tech. Specification 36.104 V 14.0.0, Jun. 2016.